\documentclass[conference]{IEEEtran}
\IEEEoverridecommandlockouts
\usepackage{cite}
\usepackage{amsmath,amssymb,amsfonts}
\usepackage{algorithmic}
\usepackage{graphicx}
\usepackage{textcomp}
\usepackage{xcolor}
\usepackage{subfigure}
\usepackage{multirow}
\usepackage{url}
\usepackage{booktabs}
\def\BibTeX{{\rm B\kern-.05em{\sc i\kern-.025em b}\kern-.08em
    T\kern-.1667em\lower.7ex\hbox{E}\kern-.125emX}}
\begin{document}

\title{Demo: Untethered Haptic Teleoperation for Nuclear Decommissioning using a Low-Power Wireless Control Technology }

\author{
\IEEEauthorblockN{Joseph Bolarinwa\(^*\), Alex Smith\(^*\), Adnan Aijaz\(^\dagger\), Aleksandar Stanoev\(^\dagger\), Manuel Giuliani\(^*\)  }
\IEEEauthorblockA{
\(^*\)\text{Bristol Robotics Laboratory, University of the West of England, Bristol, United Kingdom}\\
\(^\dagger\)\text{Bristol Research and Innovation Laboratory, Toshiba Europe Ltd., Bristol, United Kingdom}\\
firstname.lastname@brl.ac.uk; firstname.lastname@toshiba-bril.com}
} 



\maketitle


\begin{abstract}
Haptic teleoperation is typically realized through wired networking technologies (e.g., Ethernet) which guarantee performance of control loops closed over the communication medium, particularly  in terms of latency, jitter, and reliability. This demonstration shows the capability of conducting haptic teleoperation over a novel low-power wireless control technology, called GALLOP, in a nuclear decommissioning use-case. It shows the viability of GALLOP for meeting latency, timeliness, and safety requirements of haptic teleoperation. Evaluation conducted as part of the demonstration reveals that GALLOP, which has been implemented over an off-the-shelf Bluetooth 5.0 chipset, can be a replacement for conventional wired TCP/IP connection, and outperforms WiFi-based wireless solution in same use-case. 
\end{abstract}

\section{Introduction}
Nuclear decommissioning is a prominent application of haptic teleoperation where robots are used due to the risk of harmful radiation exposure. Teleoperation systems for handling nuclear waste are typically tethered, i.e., based on a wired connection. However, the complexity of nuclear facilities, the risk of cable disconnection, and the emerging requirement of employing mobile robotic platforms necessitate exploration of wireless technologies for haptic teleoperation. 

Wireless technologies replacing existing wired connections in haptic teleoperation systems are expected to provide wire-like performance, especially in terms of latency, timeliness (minimal jitter), reliability, and safety. This is to guarantee the performance of real-time bi-directional exchange of command and feedback messages, between master and slave robots, which closes a control-loop over the communication network. 

In this demonstration, we show the viability of untethered haptic teleoperation using a low-power wireless control technology, called GALLOP \cite{GALLOP_journal}, which has been specifically designed for meeting the stringent requirements of wireless closed-loop control. We also show evaluation results from our teleoperation testbed for GALLOP-based operation and benchmark the performance against  wired connectivity and wireless connectivity based on Wi-Fi. To the best of our knowledge, this is one of the first technology demonstrations of wireless haptic teleoperation using a low-power wireless control technology (which is based on a Bluetooth 5.0 chipset).

\section{Demonstration Overview}
Our demonstration is based on a leader-follower haptic teleoperation setup and depicts typical tasks performed in a nuclear decommissioning use-case. The leader robot follows a human input for the required tasks (i.e., it is operated by a human). The follower robot is expected to follow a similar trajectory as the leader robot. Exchange of command and feedback messages follows a TCP/IP communication protocol. Teleoperation tasks are performed independently over conventional wired and wireless connections. We have examined key differences in performance (measured in terms of position and velocity errors, as well as responsiveness scores from \emph{heuristic} testing) between wired and wireless communication protocols.

\begin{figure}
\centering
\includegraphics[scale=0.32]{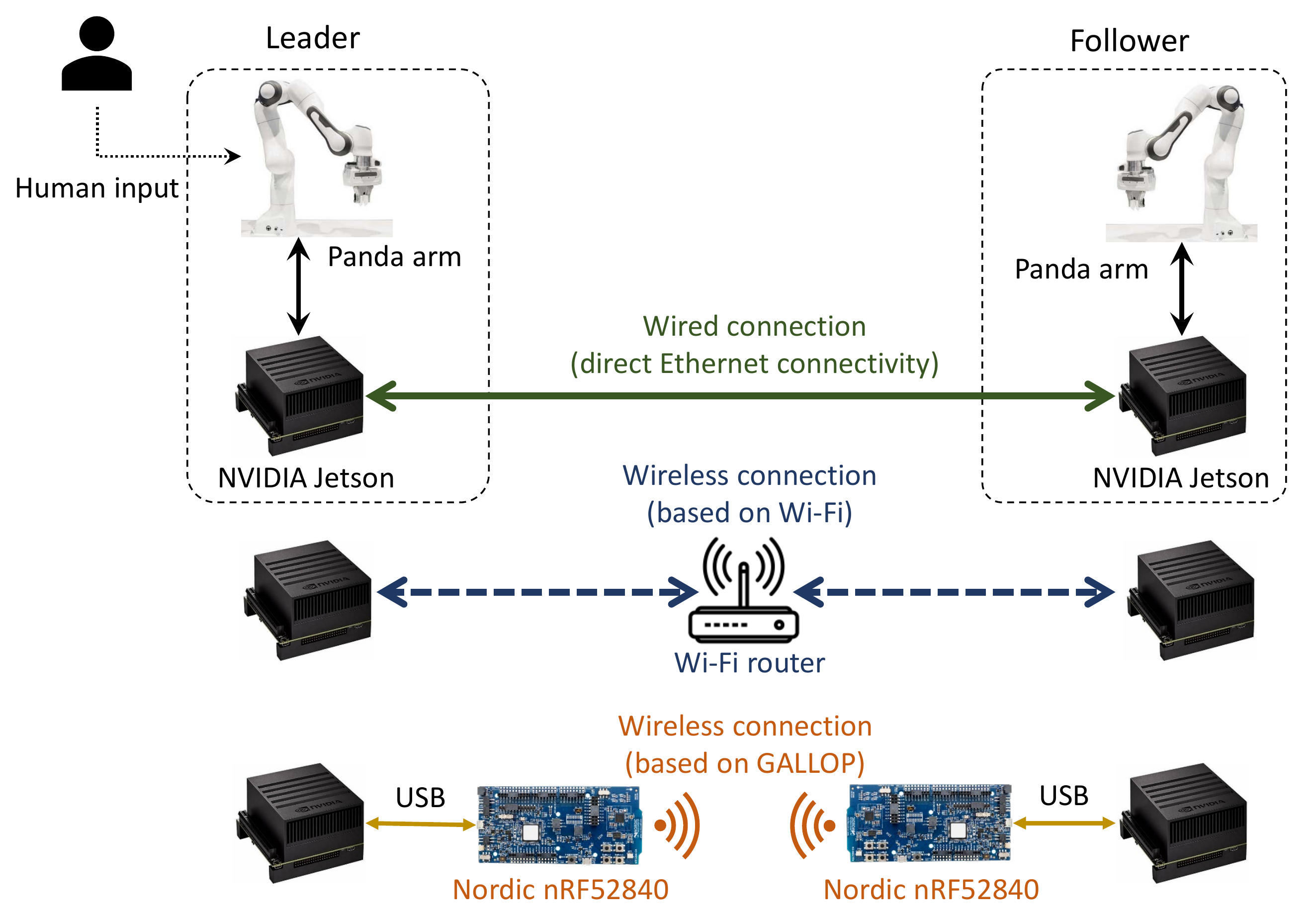}
\caption{Haptic teleoperation setup with wired and wireless connections.}
\label{demo_setup}
\vspace{-1.5em}
\end{figure}

\section{Design and Implementation}

\subsection{Teleoperation Setup}
Our demonstration setup is shown in Fig. \ref{demo_setup}. We employ Franka Emika robotic arms \cite{FrankaEmika2021} as leader and follower ends in our setup. Computations and communication between the leader and follower robots were carried using Nvidia Jetson Xavier boards \cite{Nvidia2021} connected to the controllers of the leader and follower robots. Ubuntu operating system was installed on the boards with real-time kernel to meet the 1 kHz frequency control loop requirement of the robots.

The task for the experiment consisted of using the teleoperation system to sort a random pile of six objects into three categorized boxes, with an obstacle placed between the pile and the boxes. This tasks is similar to nuclear waste sorting for decommissioning purposes. Three expert participants - familiar with the system - ran each condition (wired, wireless (Wi-Fi), and GALLOP communication) three times in a randomized order. After each experimental run they were asked to feedback on three heuristics (responsiveness, smoothness and perceived safety) based on a 5-point Likert scale.

\subsection{Wireless Control Technology}
Our demonstration employs GALLOP, which is Toshiba's proprietary technology, as a wire-replacement solution for haptic teleoperation. GALLOP has been designed for handling wireless closed-loop control with ultra-fast dynamics on the order of milliseconds (ms). It implements a control-aware schedule that handles bi-directional exchange of control and feedback messages with very low latency and zero jitter. It also implements novel techniques for achieving very high reliability in harsh wireless environments. 
In this demonstration, GALLOP handles bi-directional haptic data exchange between the leader and the follower robots. The GALLOP implementation used in this demonstration is based on a Bluetooth 5.0 wireless chipset, i.e., Nordic nRF52840 (https://www.nordicsemi.com/Products/nRF52840), and uses the 2 Mbps Physical (PHY) layer. The nRF52840 boards are directly connected to the Jetson boards over a USB interface.

\subsection{Performance Aspects}
Position and velocity errors were calculated as the difference between leader and follower arm feedback at every time step, i.e. $e(t) = q_l(t) - q_f(t)$ and $\dot e(t) =\dot q_l(t) -\dot q_f(t)$ where $e$ and $\dot e$ are joint position and velocity errors respectively, and the subscripts $\cdot _l$ and $\cdot _f$ denote feedback from the leader and follower manipulators respectively. The RMS for each joint is calculated and then summed together to get the metric $\epsilon$ for position error and $\dot \epsilon$ for velocity. 
To determine if statistically significant differences appear between wireless (Wi-Fi), wired and GALLOP conditions, the Wilcoxon Sign-Rank Test with a Bonferroni correction was used.  For position errors, the analysis shows no significant difference between the wired and wireless (Wi-Fi) conditions ($Z = -0.454$, $p = 0.65$), however there was a statistically significant reduction between GALLOP and wired ($Z = -3.18$, $p = 0.001$), and GALLOP and Wi-Fi ($Z = -2.76$, $p = 0.006$).
Examining the velocity errors $\dot \epsilon$, no significant difference was found.

\begin{table}
\centering
\caption{Summary of results for position and velocity errors}
\begin{tabular}{l|r|r|r}
& \multicolumn{3}{|c}{Position error ($\epsilon$)} \\ 
            & Wireless (Wi-Fi)     & Wired         & GALLOP \\
            \hline
Mean        & 3.00        & 3.15        & \textbf{2.38} \\
$\sigma$    & 0.390        & 0.585        & 0.393  \\
Range       & 1.12        & 2.18        & 1.34   \\
IQR         & 0.614         & 0.423         & 0.628 \\
\hline
& \multicolumn{3}{|c}{Velocity error ($\dot\epsilon$)} \\ 
 \hline 
Mean        & \textbf{1.60} & 1.76        & 1.70 \\
$\sigma$    & 0.363        & 0.411        & 0.403  \\
Range       & 1.34        & 1.39        & 1.32   \\
IQR         & 0.363         & 0.496         & 0.541 \\
\hline
\end{tabular}
\label{tab:pos-errors}
\vspace{-1.5em}
\end{table}


\begin{table}
\centering
\caption{Summary of results for heuristic questionnaires}
\begin{tabular}{l|r|r|r}
            & \multicolumn{3}{|c}{Safety} \\
            & Wireless (Wi-Fi)      & Wired         & GALLOP \\
            \hline 
Mean        & 3.53          & \textbf{4.93}         & 4.67 \\
$\sigma$    & 0.834         & 0.258        & 0.617  \\
Range       & 2             & 1             & 2   \\
IQR         & 1             & 0             & 0.750 \\
            \hline
            & \multicolumn{3}{|c}{Smoothness} \\
            & Wireless (Wi-Fi)     & Wired         & GALLOP \\
            \hline 
Mean        & 3.33          & \textbf{4.53} & 4.33 \\
$\sigma$    & 0.817        & 0.743        & 0.617  \\
Range       & 3            & 2              & 2    \\
IQR         & 0.75          & 1         & 1 \\
            \hline
            & \multicolumn{3}{|c}{Responsiveness} \\
            & Wireless (Wi-Fi)      & Wired         & GALLOP \\
            \hline 
Mean        & 3.53          & 4.73        & \textbf{4.87} \\
$\sigma$    & 0.990         & 0.458       & 0.352  \\
Range       & 3             & 1           & 1   \\
IQR         & 1             & 0.750       & 0 \\
\hline
\end{tabular}
\label{tab:heuristics}
\vspace{-1.5em}
\end{table}

\begin{figure}[htbp]
\centerline{\includegraphics[width=0.7\columnwidth]{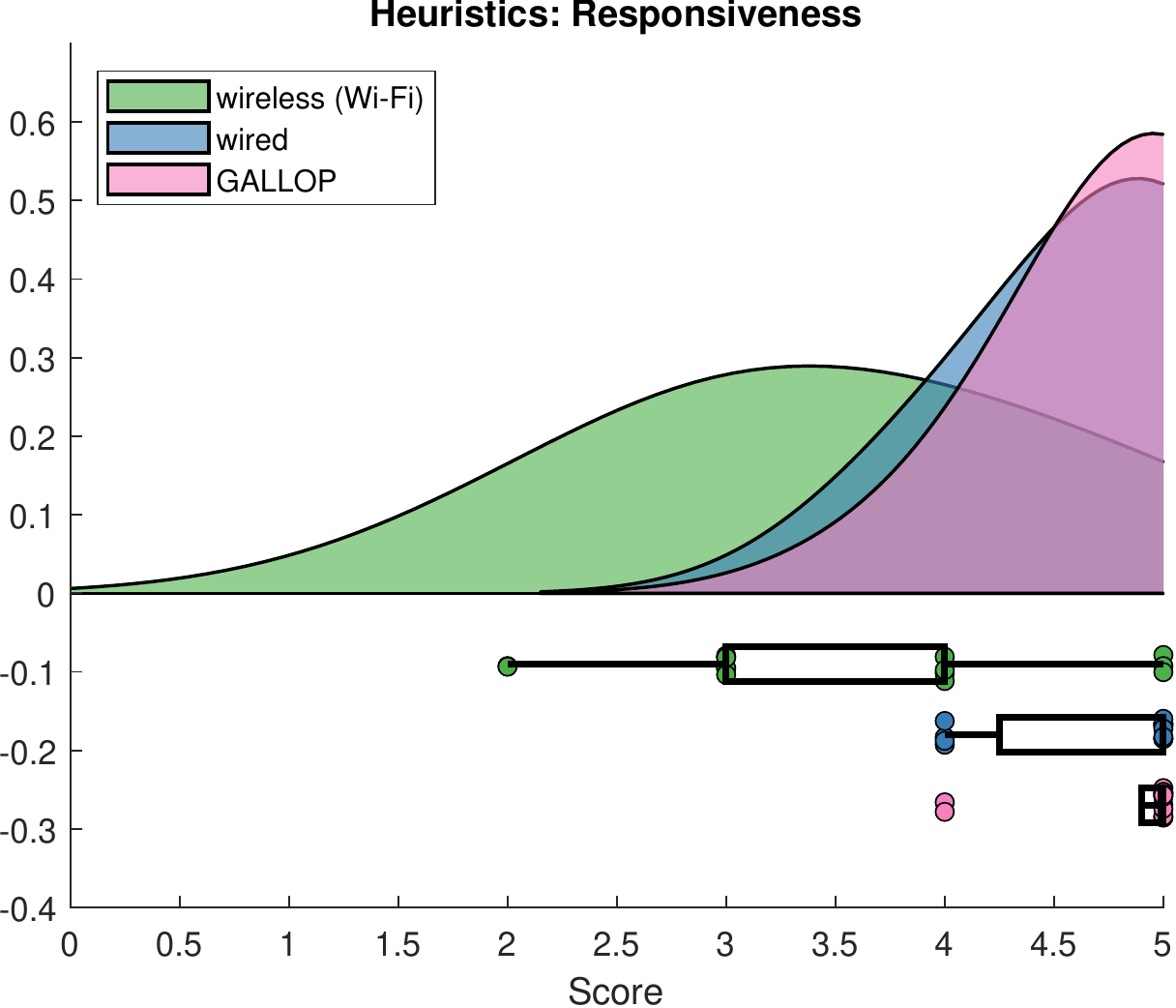}}
\caption{Responsiveness scores from Heuristic testing.}
\label{fig:responsive}
\vspace{-1.5em}
\end{figure}

Examining the results of the heuristics, for all criteria (safety, smoothness and responsiveness) there was no significant result between wired and GALLOP connections, but significance was found between wired and Wi-Fi and GALLOP and Wi-Fi connections. This can be seen reflected in \ref{fig:responsive} results for responsiveness.

\section{Remarks}
This demonstration reveals that GALLOP can be a suitable low-power (and low-cost) cable replacement solution for haptic teleoperation. It meets the key requirements of robustness, reliability, and responsiveness, and provides safe and stable teleoperation which is particularly important for the nuclear industry. A video of the demonstration is available at https://www.youtube.com/watch?v=oaqA9sn5n-I

\bibliographystyle{IEEEtran}
\bibliography{Tele_bib.bib}
\end{document}